\documentclass[twocolumn,showpacs,nofootinbib]{revtex4}

\def\be{\begin{equation}}
\def\ee{\end{equation}}
\def\bea{\begin{eqnarray}}
\def\eea{\end{eqnarray}}
\input epsf

\begin{document}

\title{{\LARGE Post-inflationary behavior of adiabatic perturbations\\ and tensor-to-scalar ratio}}
\author{Andrei Linde$^{1,2}$, Viatcheslav Mukhanov$^3$ and Misao Sasaki$^2$}
\affiliation{$^1$Physics Department, Stanford University, Stanford CA 94305-4060, USA\\
$^2$Kyoto University, Yukawa Institute, Kyoto, 606-8502, Japan\\
$^3$Sektion Physik, LMU, Theresienstr. 37, Munich, Germany}

\begin{abstract}
We explain why it is so difficult and perhaps even impossible to increase
the cosmological tensor-to-scalar perturbation ratio during the
post-inflationary evolution of the universe. Nevertheless, contrary to some
recent claims, tensor perturbations can be relatively large in the simplest
inflationary models which do not violate any rules of modern quantum
field theory.
\end{abstract}

\pacs{98.80.Cq \hskip 3.3 cm YITP-05-41 \hskip 3.3 cm \ astro-ph/0509015}

\maketitle


\section{Introduction}

Tensor perturbations (gravitational waves)~\cite{Starobinsky:1979ty}
produced during inflation~\cite{book,LL,mukhbook} may serve as an important
tool helping us to discriminate among different types of inflationary models.
These perturbations typically give a much smaller contribution to the CMB
anisotropy than the inflationary adiabatic scalar perturbations~\cite{mukh1,mukh2,kodsas}.
Nevertheless, in the simplest versions of chaotic inflation models with potentials
$\sim \phi^{n}$~\cite{Chaot} the amplitude of tensor perturbations can be
large enough to be detected.
Meanwhile in most of the other models, such as new inflation~\cite{New} and
hybrid inflation~\cite{Hybrid}, tensor perturbations typically are too small~\cite{LL}.
Therefore a discovery of the B-mode of the CMB polarization, which is
related to tensor metric perturbations, would be a strong argument in favor
of the simplest versions of chaotic inflation, whereas the absence of the B-mode would
rule out the simplest versions of chaotic inflation without helping much in
distinguishing between many other versions of inflationary theory.

Recently it was argued, however, that the contribution of tensor perturbations to the
CMB anisotropy can be much greater than expected. One may consider the models where inflation occurs at very high energy scales, which increases the amplitude of scalar and tensor perturbations. Then one may try to find a way to significantly
suppress the amplitude of scalar perturbations due to some post-inflationary
evolution. This would boost up the relative amplitude of tensor
perturbations and make them detectable in a broad class of inflationary
models~\cite{Bartolo:2005jg,Sloth:2005yx}.

The basic idea of Refs.~\cite{Bartolo:2005jg,Sloth:2005yx} can be explained as follows.
Suppose, for example, we have a hot post-inflationary universe with energy density
$\rho \sim T^{4}$, and with density perturbations $\delta \rho \sim 4 T^{3}\delta T$.
Now let us assume, for example, that some homogeneous scalar field begins to dominate 
the evolution of the universe, and then it decays.  If the field decay occurs 
simultaneously in all parts of the universe, it increases the total energy density of 
radiation without increasing the amplitude of its perturbations. As a result, 
the ratio $\delta\rho/\rho$ decreases.

The problem with this idea is that such a decay cannot occur 
in all parts of the universe simultaneously. To be more precise, one may work in the 
gauge where the spatial scalar curvature of the universe vanishes, called
the flat slicing. In this gauge, 
the Hubble constant and density $\rho$ are functions of both space and time, 
$\rho = \rho(t,x^{i})$, $H = H(t,x^{i})$, but they are related to each other by 
the standard flat space equation $H^{2} = \rho/3$, in units $M_{p} = 1$. Scalar 
perturbations of the metric in this gauge are directly related to the non-simultaneous 
end of inflation in different parts of the universe, due to quantum fluctuations of 
the inflaton field. One may think about different parts of the  universe of a 
superhorizon size as of separate flat universes which started their post-inflationary 
expansion at slightly different moments of time determined by 
$\rho(t+\delta t,x^{i}) = \rho_{e}$, where $\rho_{e}$ is the energy density at the 
end of inflation. 
The subsequent dynamics of all physical processes in the universe on the superhorizon 
scale will be determined not by the global time $t$, but by the local values of density 
of matter, which will be a function of $t+\delta t$.

A more rigorous, mathematical description of the above is possible in terms 
of the $\delta N$ formalism, in which the amplitude of the adiabatic
perturbation generated during inflation is given by the difference in the number
of $e$-foldings relative to the background from the time of horizon
crossing during inflation on the flat hypersurface to the end time of inflation
which occurs on a comoving (or uniform density)
hypersurface~\cite{Starobinsky,Sasaki:1995aw,LythRiotto,Wands:2000dp,Lyth:2004gb,Lee:2005bb}.

Consider for example any other scalar field $\sigma$. Suppose that this
field decays after it becomes dominating the energy density of
the universe. This situation is very similar to the one encountered in the
curvaton scenario, but now we will assume,
following~\cite{Bartolo:2005jg,Sloth:2005yx}, that the fluctuations generated
in this field during inflation are negligibly small, so that they do not induce
any new adiabatic perturbations after the field $\sigma$ decays. The main assumption
made in~\cite{Bartolo:2005jg,Sloth:2005yx} was that this field begins
oscillating or decays at the same time $t$ in all parts of the universe. But
the field $\sigma$ does not ``know'' anything about the global
cosmological time $t$, it depends only on the local energy density of the universe, 
local temperature, etc. 

This means, for example, that a massive field $\sigma$ with an effective potential
$m_{\sigma}^{2}\sigma^{2}/2$ will start oscillating not at some given time $t$, 
but at the time $t+\delta t$ when the local value
of Hubble constant $H(t+\delta t)$ will become smaller than $m_{\sigma}$. At
that moment, energy density $m_{\sigma}^{2}\sigma^{2}/2$ of the homogeneous
field $\sigma$ everywhere in the universe will constitute the same fraction
of the local energy density of mater given by $3 H^{2}(t+\delta t)$, and, consequently,
the same fraction of the energy density of radiation.
As a result, the motion of the field $\sigma$, its density, and its
subsequent decay will be synchronized in the same way as all other matter in the 
universe. This is the main reason  why its decay cannot reduce the amplitude of adiabatic
perturbations; they will be simply redistributed from one type of matter to another.

For the particular models studied in \cite{Bartolo:2005jg,Sloth:2005yx}
 this negative conclusion follows directly from the standard theory of
 inflationary perturbations in the theories involving only one scalar 
field~\cite{mukh1,mukh2,kodsas}.
Indeed, the second scalar field discussed in these models behaved essentially 
like a decaying fluid; no effects related to inflationary quantum fluctuations 
induced in the second field were considered in \cite{Bartolo:2005jg,Sloth:2005yx}; 
in this respect see also comments in Ref.~\cite{Lyth:2005sh}. 

The authors of Refs.~\cite{Bartolo:2005jg,Sloth:2005yx}  also realized 
that the original models described in their papers did not provide damping of 
adiabatic perturbations. Still one may wonder whether one could propose a different, 
more sophisticated mechanism to suppress superhorizon inflationary perturbations. 
In particular, instead of the possibility to ``dilute'' the original adiabatic
perturbations as suggested in~\cite{Bartolo:2005jg,Sloth:2005yx}, one could 
speculate about the possibility to generate isocurvature perturbations~\cite{Ax},
convert them to adiabatic perturbations by the curvaton
 mechanism~\cite{LM,LW,Moroi:2001ct,Enqvist:2001zp} 
(or use the mechanism proposed in~\cite{mod1}),
and then fine-tune the amplitude and phases of these
perturbations to cancel the original adiabatic perturbations.
In other words, instead of damping the adiabatic perturbations,
one could try to cancel them by perturbations of a different type.

In this paper we will argue that there is no natural way to suppress the
original adiabatic perturbations. In particular,  adiabatic perturbations
generated by the curvaton mechanism are produced from quantum fluctuations
of the curvaton field, which are generated independently of the quantum
fluctuations of the inflaton field. Therefore even if one fine-tunes
 the {\it amplitude} of the curvaton  perturbations, they cannot
have the same {\it phase} as the initial adiabatic perturbations, 
which would be necessary for their cancellation.

Our results provide an additional demonstration of robustness of inflationary
predictions. On the other hand, one could worry that  if it is impossible to boost the tensor-to-scalar ratio,  then our chances to detect tensor perturbations are very slim. This point of view was expressed in  \cite{Bartolo:2005jg,Lyth:1996im}, where it was argued that it is difficult to construct a satisfactory model of inflation  predicting detectable tensor perturbations. In Section~\ref{Grav} we explain why we are more optimistic in this respect.

\section{Evolution of the superhorizon scalar perturbations}\label{Evolution}

The theory of  cosmological perturbations in inflationary theories involving
 one scalar field, the inflaton,
is well-developed, see, e.g., \cite{mukh2,kodsas,book,mukhbook,LL}. There
is no room for speculation there, and the conclusion about the ratio of the
gravitational waves to adiabatic perturbations is robust. Namely, the
amplitude of tensor perturbations is suppressed compared to the amplitude of 
scalar perturbations by a factor of order 
$\left( 1+p/\varepsilon\right) \ll 1$,
where the ratio of the effective pressure $p$ to the energy
density $\varepsilon $ is estimated when the mode with the corresponding
comoving wavenumber crosses the horizon during inflation. Therefore to find
out whether the ratio of the tensor and scalar perturbations can substantially 
grow due to some post-inflationary physical processes one must consider  
multi-component models, where different scalar fields play important roles
at different stages of evolution. For instance one could consider the
situation when after inflation (due to the inflaton), another scalar field
(for example, curvaton), which was subdominant during inflation, takes over
and begins to dominate. Assuming that this second scalar field was initially
distributed nearly homogeneously, one may wonder whether the inflationary
metric fluctuations  can be suppressed after this  initially homogeneous
scalar begins to dominate. We will argue below that the
original inflationary perturbations will be generically transferred to the second
scalar field and hence  the resulting fluctuations will be at least
as large as the original inflationary fluctuations.

\subsection{Isentropic fluids}

In order to make our conclusions intuitively clear, we begin with the
case of isentropic ideal hydrodynamical fluids, where one can find an explicit
solution for superhorizon modes even in multicomponent media.
It is very instructive to find out how the adiabatic perturbations 
generated in one of these fluids become transferred to other liquids even 
if they interact only gravitationally.

If different ideal fluids interact only gravitationally then the
conservation law 
\begin{equation}
T^{\alpha}_{~\beta ;\alpha }=0  \label{5}
\end{equation}%
is valid for every component separately. Taking into account that the
off-diagonal spatial components of the energy-momentum tensor vanish the
metric of a flat universe with small scalar perturbations can be written as
(see, for example, \cite{mukhbook})%
\begin{equation}
ds^{2}=a^{2}\left( \eta \right) \left[ \left( 1+2\Phi \right) d\eta
^{2}-\left( 1-2\Phi \right) \delta _{ik}dx^{i}dx^{k}\right] .  \label{6}
\end{equation}%
This is called the longitudinal or Newton gauge. In what follows,
perturbation variables are those defined in this gauge unless otherwise
stated.
Then for $each$ $\ $fluid the linearized conservation law 
$T^{\alpha}_{~0;\alpha}=0$ takes the following form:
\begin{equation}
\delta \varepsilon _{J}^{\prime }+3\mathcal{H}\left( \delta \varepsilon
_{J}+\delta p_{J}\right) -3\left( \varepsilon _{J}+p_{J}\right) \Phi
^{\prime }+a\left( \varepsilon _{J}+p_{J}\right) u_{,i}^{i}=0 .  \label{7}
\end{equation}%
Here $\mathcal{H=}a^{\prime }/a$, the prime denotes the derivative with
respect to the conformal time $\eta $, and $J$ numerates different fluids. On
the superhorizon scales, the last term in Eq.~(\ref{7}) can be neglected.
 Using equation%
\begin{equation}
\varepsilon _{J}^{\prime }=-3\mathcal{H}\left( \varepsilon _{J}+p_{J}\right),
\label{7a}
\end{equation}%
which is also valid for each fluid separately, we can rewrite (\ref{7}) as%
\begin{equation}
\left[\frac{\delta \varepsilon_{J}}{\varepsilon_{J}+p_{J}}-3\Phi\right]^{\prime}
=\frac{\varepsilon_{J}^{\prime }\delta p_{J}-p_{J}^{\prime}
\delta \varepsilon_{J}}{\left( \varepsilon_{J}+p_{J}\right) ^{2}}\,.
\label{8}
\end{equation}%
Introducing standard notations%
\begin{eqnarray}
&&\zeta _{J}
=\Phi -\frac{\delta \varepsilon _{J}}{3\left( \varepsilon_{J}+p_{J}\right)}
=\Phi +\mathcal{H}\frac{\delta \varepsilon _{J}}
{\varepsilon _{J}^{\prime }},
\nonumber\\
&&\delta p_{J}^{\rm nad}
=\delta p_{J}-\frac{p_{J}^{\prime }}{\varepsilon _{J}^{\prime }}
\delta \varepsilon _{J},
\label{9}
\end{eqnarray}%
where $\zeta_J$
describes the curvature perturbation on slices of homogeneous density
for each fluid component, one can represent Eq.~(\ref{8}) in the following form:%
\begin{equation}
\zeta _{J}^{\prime }=\frac{\mathcal{H}}{\varepsilon _{J}+p_{J}}\delta
p_{J}^{\rm nad}.  \label{10}
\end{equation}%
Note that in deriving (\ref{10}) we did not use any  Einstein equations
for perturbations. 

In the case of isentropic fluids $p_{J}=p_{J}\left( \varepsilon _{J}\right) $
we have%
\begin{equation}
\delta p_{J}^{\rm nad}=\delta p_{J}-\frac{p_{J}^{\prime }}{\varepsilon _{J}^{\prime
}}\delta \varepsilon _{J}=0  \label{11}
\end{equation}%
for each component and 
\begin{equation}
\zeta _{J}^{\prime }=0.  \label{11a}
\end{equation}%
Hence $\zeta _{J}$ is conserved for each fluid separately on superhorizon
scales. We would like to point out that when we skipped the last term in
(\ref{7}) we have lost the decaying mode of the scalar perturbations, which
anyway soon becomes negligible.

Let us introduce the \textquotedblleft master $\zeta $\textquotedblright\ 
\begin{equation}
\zeta \equiv \Phi +\frac{2}{3}\frac{\Phi ^{\prime }+\mathcal{H}\Phi }{%
(1+p/\varepsilon )\mathcal{H}} \ ,  \label{11b}
\end{equation}%
where $p$ and $\varepsilon $ are the \textit{total} pressure and the energy
density respectively. Note that this variable is equal to minus 
the curvature perturbation on comoving slices, often denoted by ${\cal R}_c$,
i.e., $\zeta=-{\cal R}_c$~\cite{Sasaki:1995aw}. 
Although the curvature perturbation on comoving slices
and the curvature perturbation on homogeneous density slices are not identical,
they coincide on superhorizon scales. Using the $(0,0)$-component of
Einstein equations%
\begin{equation}
\mathop{\Delta}\limits^{(3)}
\Phi -3\mathcal{H}\left( \Phi ^{\prime }+\mathcal{H}\Phi \right)
=4\pi Ga^{2}\left( \sum \delta \varepsilon _{J}\right) ,  \label{11c}
\end{equation}%
where one can neglect the first term for the long-wavelength perturbations,
we find%
\begin{equation}
\zeta =\frac{1}{\left( \varepsilon +p\right) }\left( \sum_{J}\left(
\varepsilon _{J}+p_{J}\right) \zeta _{J}\right) . \label{11d}
\end{equation}%
After inflation, when $\rho+p\sim O(\rho) $, the value of $\zeta $ is of
order the gravitational potential, that is, $\Phi =O\left( 1\right)\times\zeta.$
In distinction from $\zeta _{J}$, the master $\zeta $ is not in general
conserved. However, irrespective of what happens with the perturbations
 at some intermediate time scale,
 Eq.~(\ref{11d}) implies that at some final moment of time 
$\eta_{f}$, when the $J=F$ fluid dominates, 
\begin{equation}
\zeta \left( \eta _{f}\right) \simeq \zeta _{F}\left( \eta _{f}\right)
=\zeta _{F}\left( \eta _{i}\right) \ , \label{11f}
\end{equation}%
where the latter equality follows from the conservation of $\zeta _{J}.$ At
the initial moment%
\begin{equation}
\zeta _{F}\left( \eta _{i}\right) =\left( \Phi -\frac{\delta \varepsilon _{F}%
}{3\left( \varepsilon _{F}+p_{F}\right) }\right) _{\eta _{i}} \ . \label{11s}
\end{equation}%
Here the gravitational potential $\Phi \left( \eta _{i}\right) $ is mostly
due to the fluid component $J=I$, which dominates at $\eta _{i}$ and is equal
to (see, for example, \cite{mukhbook})%
\begin{equation}
\Phi \left( \eta _{i}\right) \simeq -\frac{1}{2}\frac{\delta \varepsilon _{I}%
}{\varepsilon _{I}}\left( \eta _{i}\right) .  \label{11t}
\end{equation}%
It follows from (\ref{11f}) and (\ref{11s}) that the final value of 
$\Phi\left( \eta _{f}\right) =O\left( 1\right)\times\zeta \left( \eta _{f}\right)$ is
in general never smaller than $\Phi \left( \eta _{i}\right)$. Even if the
component which finally dominates was initially distributed homogeneously, i.e.
$\delta \varepsilon_{F}\left( \eta _{i}\right) =0$, we obtain, 
nevertheless, $\Phi \left( \eta_{f}\right) \sim \Phi \left( \eta _{i}\right)$.
Only in the exceptional case when 
\begin{equation}
\Phi \left( \eta _{i}\right) \simeq -\frac{1}{2}\frac{\delta \varepsilon _{I}%
}{\varepsilon _{I}}\left( \eta _{i}\right) =\frac{\delta \varepsilon _{F}}{%
3\left( \varepsilon _{F}+p_{F}\right) }\left( \eta _{i}\right) ,  \label{11q}
\end{equation}%
the resulting metric perturbations can be strongly suppressed compared to
their initial values. However, it is 
impossible to
realize in a natural way the condition (\ref{11q}) because it requires the
coherent correlated distribution of the fluids at all wavelengths.

\subsection{Scalar fields}

Unfortunately the results obtained above cannot be directly reformulated for several
scalar fields. In fact, there is no analog of separately conserved $\zeta_{J}$
in this case even if the scalar fields interact only gravitationally.
However, one can argue in a similar way that the cancellation of the dominant
adiabatic mode is also quite unlikely in this case. We first consider for
simplicity several noninteracting scalar fields $\varphi _{J}$ with
potentials $V_{J}\left( \varphi _{J}\right)$. Then their contributions to
the total energy density and pressure are given by%
\begin{eqnarray}
\varepsilon _{J}&=&\frac{1}{2}g^{\alpha \beta }\varphi _{J,\alpha }\varphi
_{J,\beta }+V_{J}\left( \varphi _{J}\right) ,\nonumber \\
p_{J}&=&\frac{1}{2}g^{\alpha \beta }\varphi _{J,\alpha }\varphi _{J,\beta
}-V_{J}\left( \varphi _{J}\right) .\nonumber
\end{eqnarray}
For several scalar fields it is convenient to introduce%
\begin{equation}
\zeta _{J}\equiv \Phi -\frac{\delta \varepsilon _{J}}{3\left( \varepsilon
_{J}+p_{J}\right) }\times \frac{\left( 1-\delta p_{J}/\delta \varepsilon
_{J}\right) }{\left( 1-p_{J}^{\prime }/\varepsilon _{J}^{\prime }\right) }%
=\Phi +\mathcal{H}\frac{\delta \varphi _{J}}{\varphi _{J}^{\prime }}
\label{12}
\end{equation}%
instead of $\zeta _{J}$ defined in (\ref{9}),
which describes the curvature perturbation on slices comoving
with each scalar field.\footnote{%
Strictly speaking the $\zeta$ variable cannot be used to trace the behavior
of scalar perturbations through the stage of oscillation of the scalar field
because the next to leading $k^{2}$-corrections to the long-wavelength
solution for $\zeta $ become infinite when $\varphi^{\prime}$ vanishes.
The variable $v=a\varphi^{\prime }\zeta /\mathcal{H}=a\,\delta\varphi_{flat}$,
where $\delta\varphi_{flat}$ is the scalar field perturbation on flat slices, 
should be used instead (see, for example, \cite{mukhbook}). However considering
only the leading order behavior of $\zeta $ we ignore this subtlety~here.}
Using Einstein equations for perturbations one can easily verify that only for
the case of a single scalar field the $\zeta_{J}$ defined in (\ref{9}) and 
(\ref{12}), are equal; otherwise they are different. The conservation 
equation~(\ref{7}) is still valid for each scalar field separately. However, 
in this case one cannot reduce it to a useful form similar to (\ref{10}) with 
$\delta p_{J}^{\rm nad}=0.$ Therefore we consider instead the conservation of the
total energy%
\begin{equation}
\left( \sum_{J}\delta \varepsilon _{J}\right) ^{\prime }+3\mathcal{H}%
\sum_{J}\left( \delta \varepsilon _{J}+\delta p_{J}\right) -3\sum_{J}\left(
\varepsilon _{J}+p_{J}\right) \Phi ^{\prime }=0\,,  \label{13}
\end{equation}%
where we took the advantage of considering the superhorizon perturbations
and skipped the spatial derivative terms. From (\ref{11c}),
where we neglect $\mathop{\Delta}\limits^{(3)}\Phi$,
and the $(0,i)$-component of Einstein equations 
\begin{equation}
\left( \Phi ^{\prime }+\mathcal{H}\Phi \right) =4\pi Ga^{2}\sum_{J}\left(
\varepsilon _{J}+p_{J}\right) \frac{\delta \varphi _{J}}{\varphi
_{J}^{\prime }}\,,  \label{14}
\end{equation}%
we find 
\begin{equation}
\sum_{J}\delta \varepsilon _{J}=-3\mathcal{H}\sum_{J}\left( \varepsilon
_{J}+p_{J}\right) \frac{\delta \varphi _{J}}{\varphi _{J}^{\prime }}\,.
\label{15}
\end{equation}%
Substituting this expression into (\ref{13}) and using the background
equation of motion we finally obtain%
\begin{equation}
\sum_{J}\left( \varepsilon _{J}+p_{J}\right) \zeta _{J}^{\prime }=0\,.
\label{16}
\end{equation}%
Taking the derivative of the master $\zeta $ (see Eqs.~(\ref{11b}) 
and (\ref{11d}),  which are valid for scalar fields) and using 
(\ref{16}) one gets%
\begin{eqnarray}
\zeta ^{\prime }&=&\frac{1}{\left( \varepsilon +p\right) ^{2}}\sum_{J,K}\left(
\varepsilon _{J}+p_{J}\right) ^{\prime }\left( \varepsilon _{K}+p_{K}\right)
\left( \zeta _{J}-\zeta _{K}\right) \nonumber \\
&\equiv& \sum_{J,K}F_{JK}\left( \eta
\right) \left( \zeta _{J}-\zeta _{K}\right)\,.
\label{17}
\end{eqnarray}%
One can verify that after the substitution $\varepsilon _{J}+p_{J}\rightarrow
\varphi _{J}^{\prime 2}/a^{2}$ the equations above remain
valid for the interacting scalar fields. Eq.~(\ref{17}) implies 
the existence of the adiabatic mode for which%
\begin{equation}
\zeta _{1}=\zeta _{2}=...=\zeta =\mbox{const.}\,. \label{18}
\end{equation}%
This equation is, of course, not enough to determine the behavior of all 
$\zeta _{J}$ for generic initial conditions. The variables $\zeta _{J}$
satisfy rather complicated coupled system of equations~\cite{mukhander}.
In distinction from the isentropic fluids, $\zeta _{J}$ for the scalar
fields are not conserved separately.
However, assuming that initially and finally the universe is
dominated by one of the fields or by its decay products,
one can make rather generic conclusions about
the resulting value of $\Phi \sim \zeta $ based on (\ref{17}). 

If one makes the standard assumption that the inflaton field $\varphi _{I}$ 
and its decay products dominate during and after inflation, one comes to
the standard conclusion about the evolution of the amplitude of adiabatic 
perturbations. In this case, all post-inflationary processes on the 
superhorizon scale are synchronized by the original adiabatic perturbations 
of the metric and cannot be damped. Their final amplitude depends only 
on the final equation of state, but not on local processes at
some intermediate time~\cite{mukh1,mukh2,kodsas}. 

Now we will check whether this conclusion can be modified due to effects
related to other scalar fields. We will  assume that
after inflation is over, the energy density released by the inflaton field
decays faster than the energy density of some other (curvaton) 
scalar field $\varphi_{C}$ and finally this field begins to dominate. 
The question is whether the metric perturbations, which are finally due to
the curvaton field and its decay products,
can be suppressed compared to the metric fluctuations originally generated
by the inflaton. 

Let us consider the mode with the comoving wavenumber $k$
which crosses the horizon at the moment $\eta _{k}$ during inflation.
Integrating (\ref{17}), we have for $\eta >\eta _{k}$%
\begin{equation}
\zeta_{k}\left( \eta \right) 
=\zeta_{inf}\left( \eta _{k}\right) 
+\int_{\eta_{k}}^{\eta }\left( F_{IC}-F_{CI}\right)
 \left( \zeta _{I}-\zeta _{C}\right)d\eta ,  \label{19}
\end{equation}%
where $\zeta_{inf} \left( \eta _{k}\right) $ is the adiabatic mode due to the
inflaton perturbations. It is clear from the definition in  (\ref{17}) that
the functions $F$ are non-negligible only when $\varepsilon _{I}+p_{I}\sim
\varepsilon _{C}+p_{C}$. Only during this time interval $\zeta $ is not
conserved. Let us assume that initially at $\eta =\eta _{k}$ there exists
also an isocurvature mode in addition to the adiabatic mode, so that%
\[
S_{IC}\left( \eta \right) \equiv \zeta _{I}-\zeta _{C}=f\left( \eta \right)
S_{IC}\left( \eta _{k}\right) ,
\]%
where $f\left( \eta \right) $ does not depend on $k.$ Then at some late
moment of time $\eta _{f}$ when the curvaton dominates and $\zeta $ is conserved,
we obtain%
\begin{equation}
\zeta_{k}\left( \eta _{f}\right) =\zeta_{inf}\left( \eta _{k}\right) +\left(
\int \left( F_{IC}-F_{CI}\right) f\left( \eta \right) d\eta \right)
S_{IC}\left( \eta _{k}\right) ,  \label{20}
\end{equation}%
where the integral is some $k-$independent constant. It is clear that 
$\zeta_{k}\left( \eta _{f}\right) $ can be substantially reduced only if the
second term in (\ref{20}) compensates the first one with high
accuracy. This means that the isocurvature mode should be initially highly
correlated with the adiabatic mode, the situation which we believe is not
only contrived but in fact impossible. Indeed, the fluctuations of the
inflaton and the curvaton field occur independently. These fluctuations 
have quantum origin, they represent two independent degrees of freedom,
and their phases are random. As a result, they
cannot cancel each other. Instead of that, the square of the amplitude of a
combination of these two types of oscillations will be given by the sum of
the squares of the amplitudes of both types. In other words, the curvaton
contribution can only increase the total amplitude of adiabatic
perturbations. Generically the final amplitude is determined by the
largest of the two terms in (\ref{20}). If the first term dominates, 
which is the simplest and the most general possibility,
then the final perturbations are mainly determined by the inflaton.
Otherwise, when the second term dominates, they are due to the curvaton mechanism. 

\section{Can we detect tensor perturbations?}\label{Grav}

Now we are going to analyze the argument of \cite{Lyth:1996im,Bartolo:2005jg}
against the theories where the amplitude of tensor perturbations can be
large. This argument was based on a correct observation that the amplitude
of tensor perturbations can be large only in the theories where inflation
occurs at $\phi \gtrsim M_{p}$~\cite{Lyth:1996im,Efstathiou:2005tq}. 
But then the authors of \cite{Lyth:1996im,Bartolo:2005jg,LythRiotto} made an
assumption that 
the generic expression for the effective potential can be cast in the form 
\begin{equation}  \label{LythRiotto}
V(\phi) = V_0 + {\frac{m^2}{2}} \phi^2 + {\frac{\lambda}{4}} \phi^4 + \sum_n
\lambda_n {\frac{\phi^{4+n}}{{M_p}^n}}\, ,
\end{equation}
and then they assumed that generically $\lambda_n = O(1)$, see e.g. Eq. (6)
in \cite{Lyth:1996im} or Eq. (128) in \cite{LythRiotto}. This would imply
that the behavior of $V(\phi)$ at $\phi > M_p$ is out of control.

Here we have written $M_p = 1/\sqrt{8\pi G}$ explicitly, to expose the 
implicit assumption made in \cite{Lyth:1996im,LythRiotto}.
Why do we write $M_p$ in the denominator, instead of $1000 M_p$? 
An intuitive reason is that quantum gravity is non-renormalizable,
so one should introduce a cut-off at momenta
 $k \sim M_p$~\cite{Lyth:1996im,LythRiotto}. This is a reasonable assumption,
but it does not imply the validity of Eq.~(\ref{LythRiotto}). Indeed, the
constant part of the scalar field by itself does not have any physical
meaning. It appears in Feynman diagrams not directly, but only via its
effective potential $V(\phi)$ and the masses of particles interacting with
the scalar field $\phi$. As a result, the terms induced by quantum gravity
instead of the dangerous factors ${\frac{\phi^n }{{M_p}^n}}$ contain 
the factors $\frac{V(\phi)}{{M_p}^4}$ and $\frac{m^2(\phi)}{{M_p}^2}$~\cite{book}.
Consequently, quantum gravity corrections to $V(\phi)$ become large not at
$\phi > M_p$, as one could infer from (\ref{LythRiotto}), but only at
super-Planckian energy density, or for super-Planckian masses. 

For example,
quantum gravity corrections to the effective potential in the simplest
chaotic inflation model with $V(\phi) = {\frac{m^2}{2}} \phi^2 $ 
contain such terms as $m^{2}\phi^{2}\times {\frac{m^{2}}{M_{p}^{2}}}$ 
and $m^{2}\phi^{2}\times {\frac{m^{2}\phi^{2}}{M_{p}^{4}}}$. 
Only the last term could
be dangerous, but not until $V(\phi)$ becomes greater than the Planck energy
density. This justifies chaotic inflation models in the context of the
simplest models of scalar field coupled to gravity~\cite{book}.

One way to understand our main argument is to consider the case
where the potential of the field $\phi$ is a constant, $V=V_0$, and the
field $\phi$ does not give masses to any fields. Then the theory has a 
\textit{shift symmetry}, $\phi \to \phi +c$. This symmetry is not broken by
perturbative quantum gravity corrections, so no such terms as $\sum_n
\lambda_n {\frac{\phi^{4+n}}{{M_p}^n}}$ are generated. This symmetry might
be broken by nonperturbative quantum gravity effects (wormholes? virtual
black holes?), but such effects, even if they exist, can be made
exponentially small~\cite{Kallosh:1995hi}.

However, in some theories the scalar field $\phi$ itself may have physical
(geometric) meaning, which may constrain the possible values of the fields
during inflation. The most important example is given by $N = 1$
supergravity. The effective potential of the complex scalar field $\Phi$ in
supergravity is given by the well-known expression (in units $M_p = 1$): 
\begin{equation}  \label{superpot}
V = e^{K} \left[K_{\Phi\bar\Phi}^{-1}\, |D_\Phi W|^2 -3|W|^2\right].
\end{equation}
Here $W(\Phi)$ is the superpotential, $\Phi$ denotes the scalar component of
the superfield $\Phi$; 
$D_\Phi W= {\frac{\partial W}{\partial \Phi}} + {\frac{\partial K}{\partial \Phi}} W$.
The kinetic term of the scalar field
is given by $K_{\Phi\bar\Phi}\, \partial_\mu \Phi \partial _\mu \bar\Phi$.
The standard textbook choice of the K\"ahler potential corresponding to the
canonically normalized fields $\Phi$ and $\bar\Phi$ is $K = \Phi\bar\Phi$,
so that $K_{\Phi\bar\Phi}=1$.
This immediately reveals a problem: At $\Phi > 1$ the potential is extremely
steep. It blows up as $e^{|\Phi|^2}$, which makes it very difficult to
realize chaotic inflation in supergravity at
 $\phi \equiv \sqrt 2|\Phi| > 1$.\footnote{Note, however, that this issue is not
specific to the super-Planckian values of the fields, as it leads to problems
in realizing inflation in supergravity even for $|\Phi| \ll 1$.
Thus this problem is not directly linked to detectability of tensor modes.}

It took almost 20 years to find a natural realization of chaotic inflation
model in supergravity. Kawasaki, Yamaguchi and Yanagida suggested to take
the K\"ahler potential 
\begin{equation}
K = {\frac{1}{2}}(\Phi+\bar\Phi)^2 +X\bar X
\end{equation}
of the fields $\Phi$ and $X$, with the superpotential $m\Phi X$~\cite{jap}.

At the first glance, this K\"ahler potential may seem somewhat unusual.
However, it can be obtained from the standard K\"ahler potential 
$K = \Phi\bar\Phi +X\bar X$ by adding terms $\Phi^2/2+\bar\Phi^2/2$,
which do not give any contribution to the kinetic term of the scalar fields 
$K_{\Phi\bar\Phi}\, \partial_\mu \Phi \partial _\mu \bar\Phi$. In other
words, the new K\"ahler potential, just as the old one, leads to canonical
kinetic terms for the fields $\Phi$ and $X$, so it is as simple and
legitimate as the standard textbook K\"ahler potential. However, instead of
the U(1) symmetry with respect to rotation of the field $\Phi$ in the
complex plane, the new K\"ahler potential has a \textit{shift symmetry}; it
does not depend on the imaginary part of the field $\Phi$. The shift
symmetry is broken only by the superpotential.

This leads to a profound change of the potential (\ref{superpot}): the
dangerous term $e^K$ continues growing exponentially in the directions 
$(\Phi+\bar\Phi)$ and $|X|$, but it remains constant in the direction $(\Phi-\bar\Phi)$ (shift symmetry).
Decomposing the complex field $\Phi$ into two real scalar fields,
$\Phi = {\frac{1}{\sqrt 2}} (\eta +i\phi)$, one can find the resulting
potential $V(\phi,\eta,X)$ for $\eta, |X| \ll 1$: 
\begin{equation}  \label{superpot1}
V = {\frac{m^2}{2}} \left(\phi^2 (1 + \eta^2) + |X|^2 +{\frac{3}{2}}
\eta^{2}\right).
\end{equation}
This potential has a deep valley, with a minimum at $\eta = X =0$. Therefore
the fields $\eta$ and $X$ rapidly fall down towards $\eta = X =0$, after
which the potential for the field $\phi$ becomes $V = {\frac{m^2}{2}} \phi^2$.
One can show that the inflaton potential $V = {\frac{m^2}{2}} \phi^2$ is
not modified by radiative corrections for $V \ll M_{p}^{4}$. 
This provides a very simple realization of chaotic inflation in
supergravity~\cite{jap}, and a counterexample to the arguments of
Refs.~\cite{Lyth:1996im,LythRiotto}.

Note that there are several other ways to achieve inflation in supergravity, 
such as F-term and D-term hybrid inflation models~\cite{F,D}. However,
in our opinion, the model of chaotic inflation proposed in \cite{jap} 
is by far the simplest and the most natural inflationary model based on
supergravity. It is especially interesting therefore that tensor perturbations
predicted by this model have a relatively large amplitude. Other  models of chaotic inflation in supergravity based on the idea of shift symmetry can be found in \cite{Goncharov:1985ka,Brax:2005jv}.

In string theory the situation is more complicated. A thorough investigation
of inflation in string theory became possible only very recently, after the
discovery of the KKLT mechanism of stabilization of extra dimensions~\cite{KKLT}.
As a result, only few examples of string theory inflation are presently
available. Most of these models are based on various versions of hybrid inflation,
and the amplitude of tensor perturbations there typically is very small,
see e.g. \cite{KKLMMT,D3D7}. 

There is a family of string theory motivated models where inflation is 
supposed to occur due to rolling in the axionic direction~\cite{supnat,Nflation},
as in the `natural inflation' scenario~\cite{nat}.
One should note that the combined potential of the axion field and the
volume modulus in the KKLT scenario is very much different from the effective
potential of `natural inflation'. In particular, the volume modulus is usually destabilized
near the maximum of the axion potential, where one would naively expect to have a better
chance of development of inflation. The only fully analyzed model of this
type leads to a very small magnitude of tensor perturbations~\cite{racetrack}. This model belongs to the general class of the moduli inflation models in string theory, as opposite to brane inflation \cite{KKLMMT,D3D7}. A recent version of the moduli inflation based on a modified KKLT mechanism \cite{Conlon:2005jm} also leads to small tensor perturbations.

However, it was argued in \cite{Nflation} that if one considers 
simultaneous evolution of many types of axion fields, inflation may occur in
a vicinity of the minimum of the axion potential, where the potential is
quadratic, as in the simplest chaotic inflation scenario.
This again leads to a large amplitude of tensor perturbations.
There are some other string inflation models where tensor perturbations
could be quite large if one takes into account nontrivial kinetic terms for
the inflaton field~\cite{SilvTong}.

One of the problems with all of these scenarios is that the large energy density,
which could drive inflation and produce large tensor perturbations,
tends to destabilize vacuum and decompactify extra dimensions, 
making our universe ten-dimensional \cite{Kallosh:2004yh}.
According to \cite{Kallosh:2004yh}, there is a general upper bound on 
the Hubble constant in all inflationary models based on the original 
version of the KKLT scenario: $H \lesssim m_{{3/2}}$, where $m_{3/2}$ 
is the gravitino mass. In this case the amplitude of tensor perturbations will be 
\begin{equation}
\delta_{T} \sim {H\over M_{p}} \lesssim {m_{3/2}\over M_{p}} \ .
\end{equation}
If one substitutes there the often discussed value 
$m_{3/2} \sim 1$ TeV $\sim 10^{-15}M_{p}$, one finds an extremely small 
amplitude of tensor perturbations  $ \delta_{T} \lesssim 10^{-15}$, 
which corresponds to the tensor-to-scalar ratio $r \ll 10^{-20}$. 
This is far below the most optimistic bounds for detectability of
tensor modes  \cite{Efstathiou:2005tq,Verde:2005ff,Amarie:2005in}.

However, recently a new generation of  phenomenological models was proposed,
where the gravitino mass can be very large,
see e.g. \cite{DeWolfe:2002nn,Arkani-Hamed:2004yi,Balasubramanian:2005zx}.
Also, in  a modified version of the KKLT scenario proposed in \cite{Kallosh:2004yh}
one can have inflation with the Hubble constant much higher than the gravitino mass.
This may allow stringy inflation with a  large amplitude of tensor perturbations.

\section{Conclusions}
In this paper we analyzed the possibility to suppress the amplitude of
adiabatic perturbations produced during inflation. We considered two 
possible mechanisms of such suppression: damping of initial perturbations
 by the late-time entropy release~\cite{Bartolo:2005jg,Sloth:2005yx} and 
cancellation of initial adiabatic perturbations by adiabatic perturbations 
produced, e.g., by the curvaton mechanism~\cite{LM,LW,Moroi:2001ct,Enqvist:2001zp,mod1}. We conclude that neither of these 
mechanisms can lead to suppression of the initial adiabatic perturbations.

We pointed out, however, that there are no firm theoretical arguments 
against detectability of tensor modes in inflationary theory. If tensor 
modes are detected by the future CMB polarization experiments, it will be
a great triumph and a powerful evidence in favor of the simplest versions 
of chaotic inflation. On the other hand,  in most of the other
inflationary models the amplitude of tensor perturbations is undetectably small,
and it cannot be boosted up by damping the amplitude of adiabatic perturbations
and increasing the tensor-to-scalar ratio. Therefore in planning the future
CMB polarization experiments one should keep in mind that if the amplitude of
tensor modes happens to be too small, it will rule out the simplest versions of
chaotic inflation, but it may not tell us much about many other inflationary models.

 \begin{acknowledgments}
This paper was initiated by discussions during the conference 
``The Next Chapter in Einstein's Legacy'' at the Yukawa Institute, Kyoto, 
and the subsequent workshop.  A.L. and V.M. are grateful 
to the organizers of this conference for hospitality.  The work by A.L.
 was supported by NSF grant PHY-0244728 and by Kyoto University. 
The work by M.S. was  supported in part by Monbukagakusho Grant-in-Aid for Scientific 
Research(S) No. 14102004 and (B) No.~17340075.
 \end{acknowledgments}

\end{document}